\documentclass[prd,twocolumn,groupedaddress,nofootinbib]{revtex4}
\usepackage{graphicx}
\usepackage{amsmath}
\usepackage{epsfig}

\begin{document}

\title{$B-L$ Conserved Baryogenesis}

\author{Pei-Hong Gu$^{1}_{}$}
\email{pgu@ictp.it}

\author{Utpal Sarkar$^{2}_{}$}
\email{utpal@prl.res.in}

\affiliation{$^{1}_{}$The Abdus Salam International Centre for
Theoretical Physics, Strada Costiera 11, 34014 Trieste, Italy\\
$^{2}_{}$Physical Research Laboratory, Ahmedabad 380009, India}

\begin{abstract}

In the presence of anomaly induced sphaleron process, only a $B-L$
asymmetry can be partially converted to the baryon asymmetry while
any $B+L$ asymmetry would be completely erased. Thus in any
successful baryogenesis theories, $B-L$ is usually violated
above the electroweak scale to
explain the observed matter-antimatter asymmetry of the universe.
However, if any lepton asymmetry is not affected by the sphaleron
processes, a $B-L$ conserved theory can still realize the
baryogenesis. We present here an SU(5) GUT realization of
this scenario, which naturally accommodates small masses of Dirac
neutrinos.

\end{abstract}

\maketitle

It is well known \cite{thooft1976} that there is a
$\textrm{SU(2)}_L$ global anomaly in the standard model (SM), where
the baryon number and the lepton number are both violated, but their
difference $B-L$ is still conserved. This anomaly induced anomalous
$B+L$ violating process will be suppressed by quantum tunneling
probability at zero temperature, but at finite temperature this
process could become fast in the presence of a instanton-like
solution, the sphaleron \cite{krs1985}. During the period
\cite{moore2000},
\begin{eqnarray}
100\,\textrm{GeV}\,\sim\,T_{EW}^{}\,< \,T\,
<\,T_{sph}^{}\,\sim\,10^{12}_{}\,\textrm{GeV}\,,
\end{eqnarray}
this sphaleron process will be so fast as to wash out any primordial
$B+L$ asymmetry, although it will not affect any primordial $B-L$
asymmetry and will partially transfer the existing $B-L$ asymmetry
to a baryon asymmetry. Thus any successful baryogenesis theories
above the electroweak scale have to satisfy the requirement of $B-L$
violation in addition to the CP violation and the out-of-equilibrium
conditions as required by the Sakharov's conditions
\cite{sakharov1967}. However, any lepton asymmetry in some exotic
particles or the right-handed fermions may be immune to the
sphaleron effects, which may allow a baryon asymmetry even in a
$B-L$ conserved theory. In general, the right-handed fermions
interact with the left-handed fermions through the Yukawa couplings,
and hence any asymmetry in the right-handed fermions will be
transferred to the left-handed fermions before the electroweak phase
transition and they cannot be immune to the sphaleron processes. But
if some Yukawa couplings could be very small, then the asymmetry in
the right-handed fermions will not be affected by the sphalerons and
it may be possible to have a baryon asymmetry of the universe in a
$B-L$ conserved theory \cite{kuzmin1997}.

In this note we construct grand unified theories (GUTs) where this
possibility could be realized. In GUTs based on the gauge group
SO(10) or some larger groups, in which $B-L$ is a local symmetry, it
is not possible to embed this scenario. So, we shall demonstrate
this in an SU(5) GUT, in which, $B-L$ is a global symmetry. We shall
demonstrate how this scenario could be embedded in an SU(5) GUT.

In the
usual SU(5) GUT baryogenesis scenario, we have the following Yukawa
couplings,
\begin{eqnarray} \label{yukawa1} \mathcal{L}
&\supset& -\,f_{aij}^{(1)}\,\chi_{i}^{}\,\chi_{j}^{}\,
H_{a}^{}\,-\,f_{aij}^{(2)}\,
\psi_{i}^{}\,\chi_{j}^{}\,H_{a}^{\dagger}\,+\,\textrm{h.c.}\,,
\end{eqnarray}
where $\psi_{i}^{}$, $\chi_{i}^{}$ $(i=1,2,3)$ are the
$\mathbf{5^{\ast}_{}}$ and $\mathbf{10}$ fermions, respectively
while $H_{a}^{}$ $(a=1,2)$ is the $\mathbf{5}$ Higgs scalar composed
of a $\textrm{SU}(3)_{c}^{}$ triplet Higgs $h^{}_{a}$ and a SM
$\textrm{SU}(2)_{L}^{}$ doublet Higgs $\phi_{a}^{}$. The pair decays
of $(h_{a}^{}, h_{a}^{\ast})$ can produce a baryon and lepton
asymmetry \cite{yoshimura1978}. Although the sphaleron process has
no direct impact on the right-handed fermions, the baryon and lepton
asymmetry stored in the right-handed fermions will be eventually
affected because the Yukawa interactions are sufficiently strong to
rapidly convert the right-handed quantum numbers to the left-handed
ones. Thus the baryon asymmetry and the equal lepton asymmetry will
be completely washed out by the sphaleron action.

To generate the neutrino mass, we assume the presence of singlet
right-handed neutrinos in addition to the standard
$\mathbf{5^{\ast}_{}}$ and $\mathbf{10}$ fermions and then obtain
the new Yukawa couplings,
\begin{eqnarray}
\label{yukawa2} \mathcal{L} &\supset& -\,
y_{aij}^{}\,\overline{\nu_{R_i}^{}}\,\psi_{j}^{}\,H_{a}^{}\,+\,\textrm{h.c.}\,
\end{eqnarray}
with $\nu_{R_j}^{}$ $(j=1,2,3)$ being the right-handed neutrinos.
Since the right-handed neutrinos are singlets, they can have a
Majorana mass, but a natural scale for this mass is around the scale
of grand unification. This will then induce too small a Majorana
mass of the left-handed neutrinos, which is unacceptable. Moreover
this will break $B-L$ symmetry and will not allow us to embed the
proposed scenario. We shall thus forbid the Majorana masses of
right-handed neutrinos by the requirement of conserving the $B-L$
number. Thus the neutrinos will obtain their Dirac masses,
$m_{\nu}^{} =\sum_{a}^{} y^{}_{a}\langle H_{a}^{} \rangle$, once the
Higgs fields $H_{1,2}^{}$ develop their vacuum expectation values
(\textit{vev}s). For consistency, the \textit{vev}s of the Higgs
doublets or the effective Yukawa couplings have to be very small.

The pair decays of $(h_{a}^{}, h_{a}^{\ast})$ cannot generate any
$B-L$ asymmetry. If there is CP violation, then a $B-L$ asymmetry in
$\psi_i^{}$ can be generated, but an equal and opposite amount of
$B-L$ asymmetry in $\nu_{R_j}^{}$ will compensate this, resulting in
zero $B-L$ asymmetry. However, since the effective Yukawa coupling
is very small, the $B-L$ asymmetry in $\nu_{R_j}^{}$ will not be
affected by the sphaleron processes and hence remain intact, whereas
the $B-L$ asymmetry stored in $\psi_{i}^{}$ will get converted to a
baryon asymmetry of the universe in the presence of the sphalerons.
Thus, the Yukawa interactions (\ref{yukawa1}) and (\ref{yukawa2})
will allow the $B-L$ asymmetry stored in the SM particles to
generate a baryon asymmetry of the universe, produced by the pair
decays of $(h_{a}^{}, h_{a}^{\ast})$ by the interference of
tree-level and one-loop diagrams since the Sakharov's conditions
\cite{sakharov1967} are available to satisfy in these processes.
This $B-L$ asymmetry should be proportional to the branching ratio
of the colored Higgs scalar $h_{a}^{}$ decaying into the multiplets
and the singlet right-handed neutrinos. But the constraint from the
experimental data \cite{pdg2006} requires the Yukawa couplings to be
extremely small for Dirac neutrinos unless we highly fine tune the
cancellation between $y_{1}^{}\langle H_{1}^{}\rangle$ and
$y_{2}^{}\langle H_{2}^{}\rangle$. Under the circumstances, the
magnitude of the essential branch ratio is so small that we can not
obtain the desired amount of $B-L$ asymmetry stored in the SM
particles to explain the observed matter-antimatter asymmetry of the
universe.

We shall now extend the SU(5) GUT with additional singlet fields and
a discrete $Z_2^{}$ symmetry. All the SM particles are even under
the $Z_2^{}$ symmetry, while the right-handed neutrinos are odd and
hence the Yukawa couplings (\ref{yukawa2}) are forbidden. We further
introduce some new fields including singlet fermions
$\xi_{L_{i},R_{i}}^{}$ $(i=1,...)$, $\mathbf{5}$ Higgs scalars
$H'^{}_{a}$ $(a=1,...)$, and a real singlet scalar $\eta$, in
addition to the right-handed neutrinos. The right-handed neutrinos
and the singlet fermions are assigned $B-L=-1$, while the new
scalars are assigned $B-L=0$. The Yukawa couplings of the usual
SU(5) baryogenesis model (\ref{yukawa1}) will have several new
terms, but we shall restrict some of these interactions by invoking
the $Z_2^{}$ symmetry. We shall assume that only the singlet
fermions are even under the $Z_2^{}$ symmetry, while all other
particles including the right-handed neutrinos are odd. We can then
write down the $B-L$ and $Z_2^{}$ invariant Lagrangian with all the
fields. We shall present here only the terms that are directly
related to the rest of our discussions:
\begin{eqnarray} \label{yukawa3} \mathcal{L}&\supset&
-\,f_{aij}^{(1)}\,\chi_{i}^{}\,\chi_{j}^{}\,
H_{a}^{}\,-\,f_{aij}^{(2)}\,
\psi_{i}^{}\,\chi_{j}^{}\,H_{a}^{\dagger}\,\nonumber\\
&&-\, y^{(1)}_{aij}\,\overline{\xi_{R_i}^{}}\, \psi_{j}^{}\,H_{a}^{}
-\,z_{ij}^{}\,\overline{\nu_{R_i}^{}}\,\xi_{L_j}^{}\,\eta\,\nonumber\\
&&-\,y^{(2)}_{aij}\,\overline{\nu_{R_i}^{}}\,\psi_{j}^{}\,H'^{}_{a}\,-\,\mu_{ab}^{}\,\eta
H'^{\dagger}_{a}\,H_{b}^{}\,\nonumber\\
&&-\,M_{\xi_{ij}}^{}\,
\overline{\xi_{R_i}^{}}\,\xi_{L_j}^{}\,+\,\textrm{h.c.}\,.
\end{eqnarray}
The charged fermion masses come from the \textit{vev}s of
$H_{a}^{}$'s. But the neutrinos remain massless, since the $B-L$
symmetry prevents any Majorana mass while the $Z_2^{}$ symmetry
prevents any Dirac mass. Only after the real scalar field $\eta$
develops a \textit{vev}, it induces a \textit{vev} to the fields
$H'^{}_{a}$, which in turn generate a small Dirac mass of the
neutrinos \cite{gh2006}. Thus after the breaking of both the
$Z_{2}^{}$ and electroweak symmetry, we will obtain an effective
neutrino Dirac mass term,
\begin{eqnarray}
\label{neutrinomass} \mathcal{L}&\supset& -\,m_{\nu_{ij}}^{}\,
\overline{\nu_{R_i}^{}}\,\nu_{L_j}^{}\,+\,\textrm{h.c.}\,
\end{eqnarray}
with $m_{\nu}^{} \equiv \sum_{a}^{} y^{eff}_{a}\langle H_{a}^{}
\rangle$, where the effective Yukawa couplings,
\begin{eqnarray}
y^{eff}_{} &=& -\,y_{}^{(1)}\,\frac{\langle \eta \rangle
}{M_{\xi}^{}}\,z\,-\, y_{}^{(2)}\,\frac{\langle \eta \rangle
}{M_{H'^{}_{}}^{2}}\,\mu_{}^{}\,,
\end{eqnarray}
can be highly suppressed by the ratio of the \textit{vev} of real
scalar over the heavy mass scales \cite{gh2006,rw1983}. In the first
term, the singlet fermion gives a see-saw contribution, while the
second term is generated by the effective $vev$ of the Higgs
$H'^{}_a$. Since both these terms are of the same order of
magnitude, any one of these terms could generate the required
neutrino masses. Thus we could include any one of the singlet
fermions or the Higgs field $H'^{}_a$ for the neutrino masses.
However, for baryogenesis we require all these fields.

We now explain how to predict the observed matter-antimatter
asymmetry in the present $B-L$ conserved model. Depending on the
values of masses and couplings, either the pair decays of
$(h_{a}^{}, h_{a}^{\ast})$ or the pair decays of $(\xi_{i}^{},
\xi_{i}^{c})$ and $(H'^{}_{a}, H'^{\ast}_{a})$ can contribute to the
final baryon and lepton asymmetry. For simplicity and illustration,
we consider two limiting cases. In the first one, we take
$M_{\xi,H'}^{}\gg M_{h}^{} $, thus the pair decays of
$(\xi_{i}^{},\xi^{c}_{i})$ and $(H'^{}_{a},H'^{\ast}_{a})$ can
produce a $B-L$ asymmetry stored in the SM particles. An equal and
opposite amount of $B-L$ asymmetry will be stored in the
right-handed neutrinos. These $B-L$ asymmetries will emerge through
the tree-level diagrams interfering with the self-energy corrections
\cite{gh2006,ars1998} and/or vertex corrections \cite{ghs2007} once
they go out of equilibrium. Since there are no other processes
violating the SM $B-L$ number, these produced $B-L$ asymmetry stored
in the SM particles can be partially converted to the baryon
asymmetry through the sphaleron process. However, the right-handed
neutrinos interact with the left-handed fields only through the
effective Yukawa couplings $y^{eff}_{}$, which is very small and
hence the $B-L$ asymmetry stored in the right-handed neutrinos are
not transferred to the SM particles, and hence, cannot take part in
sphaleron process.

We now consider another limiting case, where $M_{\xi,H'}^{}\ll
M_{h}^{} $. The pair decays of $(h_{a}^{}, h_{a}^{\ast})$ will
produce a baryon asymmetry stored in the SM quarks with an equal
lepton asymmetry stored in the pairs of $(\xi_{i}^{}, \xi^{c}_{i})$,
if $(\xi_{i}^{}, \xi^{c}_{i})$ satisfies departure from equilibrium, or
produce a pure baryon asymmetry if $(\xi_{i}^{}, \xi^{c}_{i})$ keeps
in equilibrium. Anyway we can eventually obtain a net $B-L$
asymmetry stored in the SM particles since even the lepton asymmetry
stored in the pairs of $(\xi_{i}^{}, \xi^{c}_{i})$ will be transferred
to the SM leptons and the right-handed neutrinos, proportional to
their branching ratios in the decays of the singlet fermions. Note, in
this case, we have not considered the contribution
from the out-of-equilibrium and CP-violation decays of
$(\xi_{i}^{},\xi^{c}_{i})$ and $(H'^{}_{a},H'^{\ast}_{a})$ to the
final SM lepton asymmetry for simplicity,
since we are flexible to keep the CP
conservation in these pair decays by choosing the proper CP phases.

In this note we propose a new scenario for baryogenesis before the
electroweak phase transition. Since a lepton asymmetry is immune
from the sphaleron action in our scenario, a net $B-L$ asymmetry can
be converted to the baryon asymmetry through the sphaleron process
and then explain the observed matter-antimatter asymmetry of the
universe although the total $B-L$ number is still conserved.

\end{document}